\newlength\smallfigwidth
\documentclass[twocolumn,english,aps,prb,showpacs,superscriptaddress,floats,amsmath,amssymb,floatfix]{revtex4}

\usepackage{color}
\usepackage{amsmath}
\usepackage{graphicx}
\usepackage{graphics}
\usepackage[pdf]{pstricks}
\usepackage{pdftexcmds}
\usepackage{ifpdf}
\usepackage{amssymb}
\usepackage{esint}
\usepackage{comment}

\makeatletter
\@ifundefined{textcolor}{}
{%
 \definecolor{BLACK}{gray}{0}
 \definecolor{WHITE}{gray}{1}
 \definecolor{RED}{rgb}{1,0,0}
 \definecolor{GREEN}{rgb}{0,1,0}
 \definecolor{BLUE}{rgb}{0,0,1}
 \definecolor{CYAN}{cmyk}{1,0,0,0}
 \definecolor{MAGENTA}{cmyk}{0,1,0,0}
 \definecolor{YELLOW}{cmyk}{0,0,1,0}
 }

\allowdisplaybreaks
\makeatother

\newcommand{\Sp}{\mathbf{S}}
\newcommand{\nn}{\nonumber}
\def\vr {{\bf r}}
\def\vk {{\bf k}}
\def\vkc {{\bf k}^{*}}
\def \ba {\begin{eqnarray}}
\def \ea {\end{eqnarray}}

\begin{document}

\title{Field induced multiple order-by-disorder  state selection in antiferromagnetic honeycomb bilayer lattice.}
\author{F.A. G\'omez Albarrac\'in}
\author{H.D. Rosales}
\affiliation{IFLP-CONICET. Departamento de F\'{i}sica, Universidad Nacional de La Plata, C.C. 67, 1900 La Plata, Argentina}

\date{\today}

\begin{abstract}

 In this paper we present a detailed study of the antiferromagnetic classical Heisenberg model on a bilayer honeycomb lattice  in
a highly frustrated regime in presence of a magnetic field. This study shows strong evidence of entropic order-by-disorder selection
in different sectors of the magnetization curve. {For antiferromagnetic couplings $J_1=J_x=J_p/3$, w}e
find that at low temperatures there are two different regions in the magnetization
curve selected by this mechanism with different number of soft and zero modes.
These regions present broken $Z_2$ symmetry and are separated by a not fully collinear classical plateau at $M=1/2$.
At higher temperatures, there is a crossover from the conventional paramagnet to a cooperative magnet.
Finally, we also discuss the low temperature behavior of the system for a less frustrated region, $J_1=J_x<J_p/3$.

\end{abstract}

\maketitle

\section{Introduction}
\label{sec-intro}

A magnetic system is called frustrated if local pairwise interactions between spins cannot be
satisfied simultaneously. Frustration can arise from competing interactions and/or from a
specific lattice geometry, as illustrated, {\it e.g.}, by a triangular lattice. Magnetic frustration
gives rise to an extremely rich phenomenology in both quantum and classical systems.
Quantum frustrated magnets are the main candidates for a variety of unconventional phases
and phase transitions such as spin liquids and critical points with deconfined fractional
excitations \cite{Senthil2004}. Frustration also plays an important role in (semi)classical
spin systems, for which the order-by-disorder mechanism \cite{Villain89,Shender82} is capable of
producing new unexpected types of long-range magnetic order.
For this phenomenon certain low-temperature configurations are favored not by the energy,
but by their entropy. If the entropy selection fails,
fluctuation between degenerate ground states may result in a cooperative paramagnet or
a classical spin liquid. \cite{Moessner98}.

In the past few years, a large body of experimental evidences were gathered for magnetic frustration in
Bi$_3$Mn$_4$O$_{12}$(NO$_3$) \cite{Smirnova2009}. This layered material consists of Mn$^{4+}$ ions with
$S=3/2$ arranged in honeycomb bilayers.  Because of the large negative value of the
Curie-Weiss temperature $\Theta_{CW}=-257$~K and a lack of long range order down to $T\sim 0.4$~K,
competing antiferromagnetic interactions inside honeycomb planes were initially suggested \cite{Smirnova2009}.
More recently, significance of an interlayer coupling inside bilayers was pointed out by Ganesh {\it et al}. \cite{Ganesh2011}
In addition, the first-principles calculations \cite{Kandpal2011} suggest presence of only weak
intralayer frustration but strong interlayer coupling and possible frustration between layers.

Motivated by these findings, we study here the minimal classical spin model with frustration {\it between}
layers. Our model includes a strong antiferromagnetic interlayer coupling $J_p$,
the nearest-neighbor exchange in honeycomb layers $J_1$, and a diagonal interaction $J_x$ between layers
comparable in magnitude with $J_1$.
The choice of couplings $J_x\approx J_1$ leads to extensive degeneracy of classical ground states, which has its origin
in complete mean-field decoupling of interlayer classical dimers.
Similar quantum models of $S=1/2$ dimers with frustrated interdimer coupling were
extensively studied in one and two dimensions. \cite{Gelfand91,Xian95,Weihong98,Wang00,Honecker00,Lin02,Liu08}
They exhibit a number of interesting physical properties, which include exact singlet ground states,
localized triplon modes, and fractional magnetization plateaus. Our work complements the
previous studies  by providing theoretical results in the (semi)classical limit $S\gg 1$.
Although we do not expect to obtain the complete phase diagram of Bi$_3$Mn$_4$O$_{12}$(NO$_3$), the obtained results may be relevant not only for this material but also for other
magnetic materials consisting of $S>1/2$ spin dimers, see, for example, Refs.~\onlinecite{Grenier04,Stone08,Tanaka14}.

In this classical system, we find a number of novel phases induced by an external magnetic field that are generated by the thermal order by disorder
effect. The selection varies along the magnetization curve with a plateau at $M=1/2$ separating
 regions with different behaviors.
The paper is structured as follows: In Sec.~\ref{sec-model} we present the Hamiltonian and
we study the ground state order based on the spherical approximation. This allows us
to identify regions in the parameter space where frustration
 plays a significant role. In Sec. \ref{sec-frustrated-point} we focus on the strongly frustrated case, where the Hamiltonian can be rewritten as a sum of the square of the total spin of connected plaquettes.
 We describe the different phases and  study the role of thermal fluctuations as a scenario for the appearance of a $M= 1/2$  plateau  and the multiple order-by-disorder selection.
In Sec. \ref{sec-frustrated-line} we extend the previous study for a less frustrated case. We conclude in Sec. \ref{sec-conclusions}  with a summary and discussion of our results.

%
\section{Spin model and Spherical Approximation}
\label{sec-model}
\begin{figure}[t]
\begin{centering}
\includegraphics[width=0.6\columnwidth]{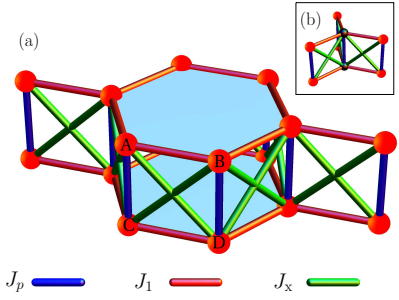}
\par\end{centering}
\caption{(Color online) (a) Frustrated honeycomb bilayer lattice. Solid spheres represent spin-S moments and the labels $A,B,C,D$
indicate the four sites of each unit cell. Intralayer nearest neighbor coupling $J_1$ and interlayers couplings $J_x$ and $J_p$ are indicated.
(b) Three plaquettes $\boxtimes$ sharing one bond. Each pair of opposite intralayer spins is shared by three plaquettes.}
\label{fig:lattice}
\end{figure}

The classical model presented here is based on the low-temperature magnetic properties of Bi$_3$Mn$_4$O$_{12}$(NO$_3$).
The main exchange interactions in this material are still disputed. \cite{Kandpal2011,Matsuda2010}
One possibility is the presence of nearest-neighbor  coupling $J_1$ and a weak second nearest-neighbor coupling $J_2$
in each layer as well as interlayer interactions $J_p$ and $J_x$. More distant interactions, such as
an intralayer third-neighbor exchange $J_3$ have also been suggested.
In our study, we consider the classical  model, where spins are replaced by unit vectors, of two non-frustrated honeycomb layers with antiferromagnetic
intralayer exchange $J_1$ coupled by frustrating antiferromagnetic $J_p$-$J_x$ bonds, see Fig.~\ref{fig:lattice}.

The general spin Hamiltonian in a magnetic field is given by
%
%
\begin{eqnarray}
\mathcal{H}=J_p\!\sum_{\vr}\left(\Sp_{\vr,A}\cdot\Sp_{\vr,C}+\Sp_{\vr,B}\cdot\Sp_{\vr,D}\right)-h\sum_{\vr,i}S^{z}_{\vr,i}  \nonumber \\
 + J_{1}\!\sum_{\langle\vr,\vr'\rangle}\left(\Sp_{\vr,A}\cdot\Sp_{\vr',B} + \Sp_{\vr,C}\cdot\Sp_{\vr',D}\right)  \nonumber \\
+J_{x}\sum_{\langle\vr,\vr'\rangle}\left(\Sp_{\vr,A}\cdot\Sp_{\vr',D} + \Sp_{\vr,B}\cdot\Sp_{\vr',C} \right)
\label{eq:Hamil}
\end{eqnarray}

\noindent where $\vr$  runs over unit cells, 
$\langle\vr,\vr'\rangle$  denotes interactions within the cell and between  nearest-neighbor cells and $i$ is the spin cell index $i={A,B,C,D}$.

This Hamiltonian has a discrete $\mathcal{Z}_2$ symmetry which corresponds to the exchange of spin pairs connected by $J_p$, i.e. 
a reflection symmetry through a plane that is perpendicular to the $J_1$ 
bonds and crosses through the middle of the plaquettes.

This symmetry will play an important role in the characterization of the 
low temperature phases and we will later discuss how this symmetry is broken in particular cases.

In the  $J_1=J_x$ case, the Hamiltonian (\ref{eq:Hamil}) has the property that can be written as a function of the total spin of the  opposite
spin pairs or ``dimers''.
We introduce  new variables  $\mathbf{P}_{\boxtimes,\alpha}=\Sp_{\boxtimes,A}+\Sp_{\boxtimes,C}$ and $\mathbf{P}_{\boxtimes,\beta}=\Sp_{\boxtimes,B}+\Sp_{\boxtimes,D}$ and performing the sum over all four-spin elementary plaquettes  $\boxtimes$ (indicated in figure \ref{fig:lattice}(b)) and we get:

\begin{eqnarray}
\mathcal{H} =  \text{C}_1 + \sum_{\boxtimes}^{N_{\boxtimes}}\left\{ \frac{J_p}{6}\left(\mathbf{P}_{\boxtimes,\alpha}^2+\mathbf{P}_{\boxtimes,\beta}^2\right) \right.\nonumber \\
+\left.J_1\left(\mathbf{P}_{\boxtimes,\alpha}\cdot\mathbf{P}_{\boxtimes,\beta}\right) -\frac{\mathbf{h}}{3}\cdot\left(\mathbf{P}_{\boxtimes,\alpha}+\mathbf{P}_{\boxtimes,\beta}\right) \right\}
\label{eq:Hpairs}
\end{eqnarray}
\noindent where the labels $\alpha,\beta$ correspond to a the spin pair $(A-C)$ and $(B-D)$ respectively; $\mathrm{h}=h\hat{z}$, $N_{\boxtimes}=\frac{3}{4}N$
is the number of plaquettes on an $N-$site lattice and $\text{C}_1$ is a constant.

Furthermore, from equation (\ref{eq:Hpairs}) it is straightforward that for the particular point $J_1=J_x=J_p/3$ (highly frustrated point)
the hamiltonian can be written, up to a constant term, as a quadratic function of the total spin of the plaquettes:
\begin{equation}
\mathcal{H}=\text{C}_2+\frac{J_p}{6}\sum_{\boxtimes}^{N_{\boxtimes}}
\Bigl(\Sp^2_{\boxtimes}-\frac{2}{J_p}{\bf h}\cdot\Sp_{\boxtimes}\Bigr),
\label{eq:Hplaquette}	
\end{equation}
where $\Sp_{\boxtimes}=\sum_{i\in \boxtimes} \Sp_i$, ${\bf h}=h\,\hat{z}$ and $\text{C}_2$ is a constant.
These two regimes $J_1=J_x$ and $J_1=J_x=J_p/3$,  where the ground state condition can be studied analytically,  will be adressed in the following sections.

As a first step  to explore the full range of parameter space, we  adopt the spherical approximation to investigate the resultant magnetic
ground state of the general Hamiltonian equation ~(\ref{eq:Hamil}) at zero magnetic field ($h=0$). In this
approximation the local fixed spin-length constraint $|\Sp_{\vr,i}|=1$ ($i=A,B,C,D$) is replaced by a global one
$\sum_{\vr}|\Sp_{\vr,i}|^2=N\,S^2/4$, where  $N$ is the number of lattice sites. \cite{Nagamiya1967}
With the  new soft constraint, the spin Hamiltonian (\ref{eq:Hamil}) can be diagonalized with
the aid of the Fourier transformation  $S^{\alpha}_{\vr}=\sum_{\vk}S^{\alpha}_{\vk}e^{i\,\vr\cdot\vk}$.
Here $\alpha=x,y,z$, and $\vr$ and $\vk$ denote the position and pseudo-momentum respectively.
The Hamiltonian then becomes

\begin{equation}
\mathcal{H}= \sum_{\vk,\alpha}\Psi^{\alpha}_{\vk}\,{\bf M}(\vk)\Psi^{\alpha}_{-\vk} \ ,
\label{Hamiltonian}
\end{equation}
where $\Psi^{\alpha}_{\vk}=(S^{\alpha}_{\vk,A},S^{\alpha}_{\vk,B},S^{\alpha}_{\vk,C},S^{\alpha}_{\vk,D})$ ($A,B,C$ and $D$ being the different sites in the unit cell as shown
in figure \ref{fig:lattice}(a)) and
\begin{equation}
{\bf M}(\vk)=\left( \begin{array}{cccc}
0 & J_1\,\gamma(\vk) & J_{p} & J_x\,\gamma(\vk)\\
J_1\,\gamma(-\vk) & 0 & J_x\,\gamma(-\vk) &J_{p}\\
J_{p} & J_x\,\gamma(\vk) & 0 &J_1\,\gamma(\vk)\\
J_x\,\gamma(-\vk) & J_{p} & J_1\,\gamma(-\vk) &0\end{array} \right)
\label{eq:matrix-Mk}
\end{equation}
The ordered ground-state energy is associated with the lowest eigenvalue
$w^{\text{min}}(\vkc)$ of the matrix ${\bf M}(\vk)$.
Four different eigenvalues of the matrix (\ref{eq:matrix-Mk}) are given by
\ba
w_{\sigma\sigma'}(\vk)=\sigma\,J_{p}+\sigma'\,|J_1+\sigma\,J_x|\,|\gamma(\vk)|
\label{eq:wk-classic}
\ea
where $\sigma,\sigma'=\pm 1$, $|\gamma(\vk)|=\sqrt{3+2\sum_{i}\cos(\vk.{\bf e_i})}$
and   ${\bf e}_1=(1,0)$, ${\bf e}_2=(1/2,\sqrt{3}/2)$, ${\bf e}_3={\bf e}_1-{\bf e}_2$.

\begin{figure}[t]
\begin{centering}
 \includegraphics[width=0.95\columnwidth]{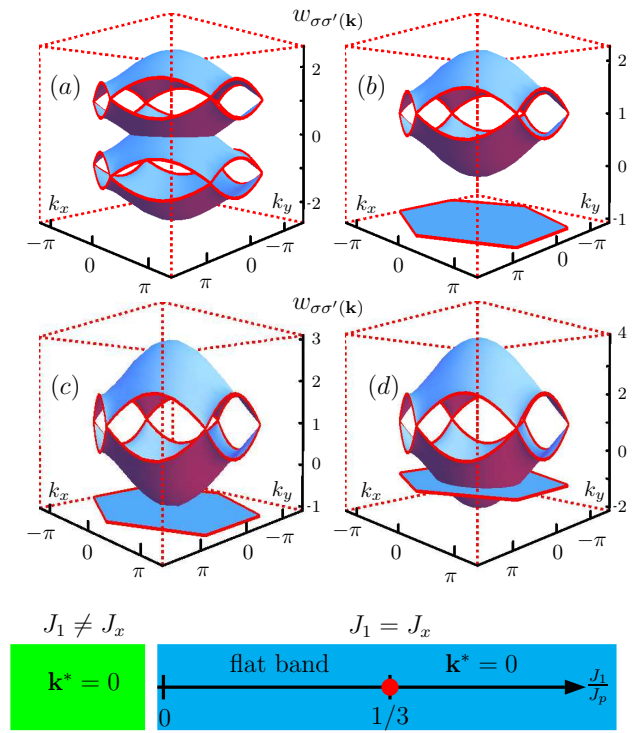}
\par\end{centering}
\caption{Dispersion relations $\omega_i(\vk)$ as a function of $\vk$ for four regions of the couplings parameter space. Panel (a) (top left)
$J_1 \neq J_x$. The lowest band has a minimum at $ \vkc = \vec{0}$. Panel (b) (top right) $J_1=J_x<J_p/3$.
The minimum are two overlapping completely flat bands, selecting no particular value of $\vk$. Panel (c) (bottom left) shows the special case of $J_1=J_x=J_p/3$. Where the minimum
has the peculiarity that at $\vkc = \vec{0}$ the two overlapping completely flat bands (that do not select a particular ordering vector) coincides with the next lowest band. Panel (d)
(bottom right) $J_1=J_x>J_p/3$ . There are two flat (overlapping) and two non-trivial bands.
The lowest band has a minimum at $\vkc = \vec{0}$.
}
\label{fig:wk-classic}
\end{figure}

Figure \ref{fig:wk-classic} shows plots of the eigenvalues from equation (\ref{eq:wk-classic})
for a few values of $\frac{J_1}{J_p},\frac{J_x}{J_p}$.
In each case, the minimum in the lowest band defines the ordering wave-vector $\vkc$.
It can be seen that for both $J_1\neq J_x$ and $J_1=J_x>J_p/3$ the ground state is unique and corresponds to
$\vk^{*}=\vec{0}$. On the other hand, for $J_1=J_x<J_p/3$ the two lowest bands are completely flat
suggesting a lack of a long-range magnetic order at zero temperature. For $J_1=J_x=J_p/3$ one of the dispersive 
upper bands  with a minimum at $\vk^{*}=\vec{0}$ touches the flat bands indicating even higher degeneracy.
 
In this situation, where $J_1=J_x\leq J_p/3$, the lowest bands $w_{--}(\vk)=w_{-+}(\vk)=-J_p$ (flat and degenerate) 
have eigenvectors $\varphi^{\alpha}_{\vk,--}=\{0,1,0,-1\}/\sqrt{2}$ and $\varphi^{\alpha}_{\vk,-+}=\{1,0,-1,0\}/\sqrt{2}$. 
So, the spherical approximation shows that in the ground state,  pairs of spins connected by $J_p$, i.e. $(A-C)$ and $(B-D)$ are 
antiferromagnetically correlated in each component ($\alpha=x,y,z$), forming ``antiferromagnetic (AF) dimers''. Therefore, because the bands are flat, the only condition 
for the ground state is that the opposite spin pairs in each cell form these AF dimers.

Thus, this simple analysis demonstrates that the magnetic system may present interesting phenomena for $J_1=J_x\le J_p/3$. 
Motivated by the spherical aproximation results and the effective Hamiltonians  (\ref{eq:Hpairs}) and (\ref{eq:Hplaquette}),
in the following sections we proceed with the investigation of the frustrated spin model (\ref{eq:Hamil}) 
in a magnetic field and at finite temperature in this particular regime $J_1=J_x$ which includes the special point $J_1=J_x=J_p/3$.

\section{Strongly frustrated point $J_1=J_x=J_p/3$}
\label{sec-frustrated-point}

Firstly, we focus or work in the highly frustrated point $J_1=J_x=J_p/3$. The corresponding Hamiltonian written as equation (\ref{eq:Hplaquette}) can
be studied minimizing the energy of each plaquette $\boxtimes$: 

\begin{equation} \label{eq:gsplaq}
\Sp_{\boxtimes}=\frac{{\bf h}}{J_p}
\end{equation}
The classical ground state is obtained when this constraint is satisfied in every block and  presents only the typical global rotation as a degeneracy.
The saturation field $h_{s}$ is determined by the condition $S^z_{\boxtimes} = 4$ which gives $h_{s}=4\,J_p$. At this value all the spins are aligned with the magnetic field.
Since the Hamiltonian can be rewritten as a sum of elementary blocks, the ground state condition (equation (\ref{eq:gsplaq})) can be fulfilled by a number of different configurations,
which leads to a highly degenerate ground state.

\begin{figure}[htb]
\begin{centering}
\includegraphics[width=0.6\columnwidth]{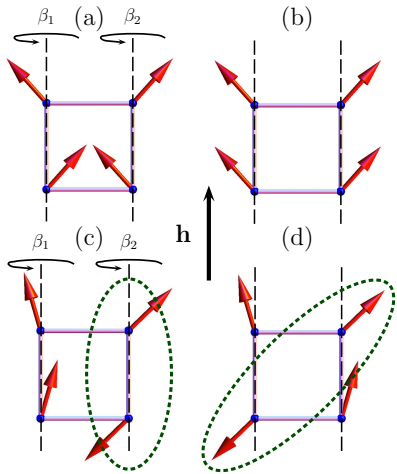}
\par\end{centering}
\caption{\label{fig:Ej-spin-conf}Examples of the spin copnfigurations with the same energy for $h/J_p < 2$ for a plaquette. In configurations
(a) and (b) all spins in the plaquette have the same projection along the magnetic field (``umbrella'' configuration). According to how the perpendicular components cancel out, given that 
plaquettes are not independent, some degeneracy is retained. In configuration (a) the azimuthal angles are free, whereas in (b) they are fixed.
In (c) and (d) the ground state condition
is fullfilled by two spins, and the other two cancel each other out. This latter pair is indicated with a dashed line. Depending on which pair cancels out, there
are still some degrees of freedom. In this case, in (c) each pair of spins has azimuthal freedom, and the polar angle in the left hand pair is arbitrary. In (d),
all the angles are fixed.
}
\end{figure}

We illustrate this degeneracy with a few examples depicted in figure \ref{fig:Ej-spin-conf}, for the case $h/J_p<2$. All the four configurations shown in the figure
satisfy the ground state condition from equation (\ref{eq:gsplaq}). Since the pairs joined by $J_p$ are shared by three plaquettes, the selection
of a particular configuration for a plaquette can fix all the system, or there might still be some degrees of freedom in each plaquette. In figure 
\ref{fig:Ej-spin-conf},
(a) and (b) are possible arrangements for an ``umbrella'' configuration, where all spins have the same projection parallel to the external magnetic field, $h$ in $z$.
The difference between (a) and (b) is simply how the azimuthal components cancel out. If these components cancel out between pairs joined by $J_1$ or $J_x$ in one
plaquette, as in (b), then all the plaquettes are fixed. However, if they cancel out between pairs joined by $J_p$, each of these pairs in each plaquette
retains the azimuthal degree of freedom. Configurations (c) and (d) are possible only for $h/J_p < 2$. The groundstate condition is fullfilled by one pair of spins 
in the plaquette, and the other two cancel each other. This particular pair that adds up to $\mathbf{S}_{\mathrm{pair}}=0$ is indicated by a dashed line.
Again, if this pair is made by two spins connected by $J_x$ (as in (d)) or by $J_1$, the configuration is fixed. If it is the $J_p$ pair, there are three 
free angles per plaquette: the two angles of the $\mathbf{S}_{\mathrm{pair}}=0$ pair and the azimuthal angle of the other one.

In the next sections we will study how the inclusion of thermal fluctuations can select specific configurations  due to a different entropy  of short-wavelength
fluctuations above degenerate configuration by the order-by-disorder mechanism at low temperatures\cite{Villain89}.

\subsection{Monte Carlo Simulations}
\label{se:MCsimulations}

In order to explore the behavior of the system at finite temperature and find whether particular phases are selected, we resorted to Monte Carlo simulations 
performed using the standard Metropolis algorithm combined with overrelaxation (microcanonical) updates\cite{Brown1987}. 
One hybrid MC step consists of one canonical MC step followed by
3-10 microcanonical random updates depending on cluster size.
Periodic boundary conditions were implemented for $N=4\times L^2$  site clusters with $L=12-72$.
At every magnetic  field or temperature  we discarded $1\times10^5$ hybrid
Monte Carlo steps (MCS) for initial relaxation and data were collected during subsequent $2\times 10^5$ MCS.

As a first step to identify different phases and corresponding transitions,  we calculate the magnetization, susceptibility and absolute value of $S^z$ defined as
\ba
M=\frac{1}{N}\sum_{\vr}S^{z}_{\vr},   &\
\chi_m={\displaystyle \frac{dM}{dh}}, &\
|S^z|=\frac{1}{N}\sum_{\vr}|S^{z}_{\vr}|.
\label{eq:MSzChi}
\ea

\begin{figure}[htb]
\begin{centering}
\includegraphics[width=0.8\columnwidth]{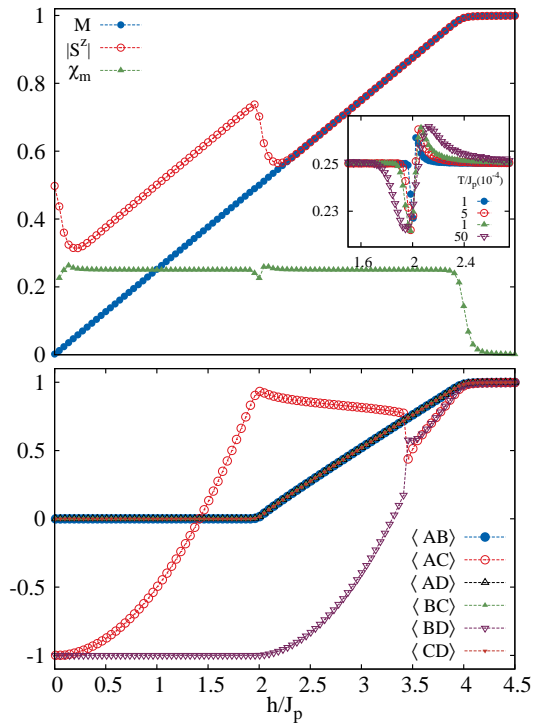}
\par\end{centering}
\caption{\label{fig:mc-MvsH} (Upper panel) Magnetization $M$, absolute value of the magnetization $|S^z|$ and magnetic susceptibility $\chi_m$ (equation (\ref{eq:MSzChi})).
The inset shows the susceptibility around $h/J_p=2$ for different temperatures.
(Lower panel)
Average of the local scalar product $\langle\Sp_{\vr,i}\cdot\Sp_{\vr,j}\rangle=\langle i j\rangle$
where $i,j$ ($i,j=A,B,C,D$ in the legend) are sites in a cell. Both plots on a $L= 48$ lattice at $T/J_p = 5\times 10^{-4}$. }
\end{figure}

In figure ~\ref{fig:mc-MvsH} we show the magnetization curve $M$, susceptibility $\chi_m$ and absolute value of $|S^z|$  as a function of the external field at temperature $T/J_p=5\times10^{-4}$.

From the top panel in figure \ref{fig:mc-MvsH} we can see that the absolute value of the magnetization $|S^z|$, which measures how "collinear" is the spin configuration, has two different behaviors.
Between  $h^*/J_p<h/J_p<2$ (where $h^*$ is a lower critical field which depends on the temperature), $|S_z|$ is shifted from total magnetization $M$ by a constant value ($1/4$),
and for  $h/J_p > 2$ they are equal. Moreover, the  susceptibility $\chi_m$ shows a dip around $h/J_p =2$ ($M=1/2$) which indicates the presence of a quasi-plateau phase.

To explore the spin configurations, we studied the average of the scalar product between two spins in an elementary plaquette $\boxtimes$,
$\langle\Sp_{\vr,i}\cdot\Sp_{\vr',j}\rangle$,
where $i,j=A,B,C,D$.
Results for $L=48$ and $T/J_p=5\times 10^{-4}$ are shown in the lower panel of figure \ref{fig:mc-MvsH}.
These results are for one  realization of the simulations. The system breaks $\mathcal{Z}_2$ symmetry, as will be discussed later.
An average of realizations will obviously restore this symmetry, as we have checked.
However, this variable is a good illustrator of the lattice symmetry breaking, and allows us to explore the spin configurations.

Figure \ref{fig:mc-MvsH} (bottom) shows that the behavior for the scalar product
of  face to face layer spins (those connected by $J_p$, $\{\Sp_{\vr,A};\Sp_{\vr,C}\}$ and $\{\Sp_{\vr,B};\Sp_{\vr,D}\}$
in figure \ref{fig:lattice}) is different than for the other ones (spins connected by intra-layer  coupling $J_1$ and diagonal interlayer coupling $J_x$).
Additionally, there is also a change in all curves at $h/J_p=2$.
This is consistent with what was discussed above for the magnetization curves in the upper panel of figure  \ref{fig:mc-MvsH}.
Below, we discuss the spin layout in each plaquette obtained from the simulations.
Configurations for different regions of magnetic field are sketched in  figure \ref{fig:spinlayout}.
All these configurations retain different degrees of freedom. This implies that along all the magnetization 
curve the ground state degeneracy is \textit{partially} lifted. These configurations and how they are selected will be discussed below.


%
\subsubsection{Low field region}
At  low temperatures and for $h^*/J_p<h/J_p<2$ (where $h^*$ is a lower critical field which depends on the temperature), the configuration is shown in
figure \ref{fig:spinlayout}(a) (notice that it has also been shown in figure \ref{fig:Ej-spin-conf} (c)).

In each plaquette, a pair of opposite spins fullfills the groundstate condition. The other pair of spins is an effective ``free spin'' or 
``antiferromagnetic (AF) dimer''
with free orientation. The behavior of the system can thus be described as two sublattices of spin pairs with different behavior. We call
this phase ``dimer phase''. This phase clearly retains some degeneracy due to the ``AF dimers''. This configuration is consistent with the behaviour 
in this region of the scalar products and the magnetization curve in figure \ref{fig:mc-MvsH}: 
in that particular case,
the pair of opposite spins that fullfills de groundstate condition is $\{\Sp_{\vr,A};\Sp_{\vr,C}\}$, and thus the scalar product grows as a function
of the magnetic field. The other opposite pair, $\{\Sp_{\vr,B};\Sp_{\vr,D}\}$, forms a ``free pair'', and therefore their scalar product is -1. The remaining
products shown in figure \ref{fig:mc-MvsH} average out to 0 because the sublattice of ``AF dimers'' is at random angles. This also explains
why in the low field region there is a $1/4$ difference between  $|S^z|$ and $M$: because of the free orientation of the ``AF dimers'', 
each spin of this kind of pair   has $\langle\Sp_{\vr,i}\rangle=0$ and $\langle|\Sp_{\vr,i}|\rangle$=0.5.
This configuration is only possible for $h/J_p < 2$, since for $h/J_p > 2$
the ground state constraint $\Sp_{\boxtimes}=\frac{{\bf h}}{J_p}$ would not be satisfied.

\subsubsection{Pseudo-plateau}
For the special value of an external field $h/J_p=2$, which is precisely one half of the saturation field $h_s$,
a weak magnetization plateau at $M=1/2$ emerges on the magnetization curve shown in Fig.~\ref{fig:mc-MvsH}.
 In the corresponding spin configuration, two spins in one vertical dimer,
{\it e.g.}, $\Sp_{\vr,A}$ and $\Sp_{\vr,C}$ are completely aligned with $h$, whereas the remaining spins
$\Sp_{\vr,B}$ and  $\Sp_{\vr,D}$ in the neighboring dimer are anti-parallel to each other, but
have, otherwise, a random orientation from cell to cell in the honeycomb layers, see Fig.~\ref{fig:spinlayout}(b).
This is a special case of the configuration shown in Fig.~\ref{fig:spinlayout}(a), with $\alpha_1=\beta_1=0$.

Notice that this is not a flat plateau, hence the name pseudo-plateau.
This is a feature that arises with temperature, since the state selection is not an energetic one \cite{Moliner2009}.
This pseudo-plateau can be defined as the region between the dip and the peak in the susceptibility. 
It is very narrow at low temperatures, then broadens and is destroyed at high temperatures.
The inset in figure \ref{fig:mc-MvsH} shows the susceptibility around $h/J_p=2$ for different values of $T/J_p$.

A remarkable property of the spin configuration at the plateau  is that it is not fully collinear.
This is in a strong contrast to the fully collinear classical plateau states in the triangular-lattice \cite{Kawamura1984}
and other frustrated spin models \cite{Zhitomirsky2000,Zhitomirsky2002}.
The root of this difference lies in a very large  configurational degeneracy of the system of frustrated classical spin dimers.
Alternation of polarized and anti-parallel spin pairs in the plateau phase is somewhat similar to the
magnetic structure of the 1/2 plateau found for the frustrated spin-1/2 ladder. \cite{Honecker00}
The latter consists of a periodic structure of triplet and singlet rungs, whereas in our case the alternation
breaks only the $\mathcal{Z}_2$ symmetry between two sites in a unit cell of the honeycomb lattice.

\subsubsection{High field region}

\begin{figure}[htb]
\begin{centering}
\includegraphics[width=1\columnwidth]{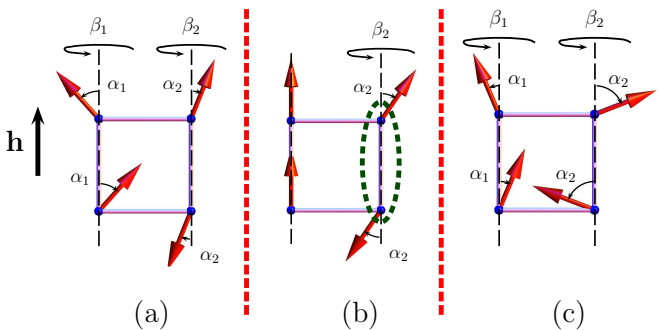}
\par\end{centering}
\caption{\label{fig:spinlayout} Spin configurations in each plaquette.
(a) $h/J_p<2$: a pair of spins at opposite sites between layers satisfies $\Sp_{\vr,A}+\Sp_{\vr,C}= \mathbf{h}/J_p$, 
and the other pair is always anti-parallel at a random position  (``free spin'' or ``AF dimer'').
(b)  $h/J_p=2$ pseudo-plateau phase: a particular case of (a), where two spins  are aligned with the magnetic field
and the other pair forms a ``free spin'' (enclosed by the dashed line)
(c)  $h/J_p > 2$: spins of sites connected by $J_{p}$ have the same $z$ projection, and opposite $xy$ components.}
\end{figure}

For $2<h/J_p<h^{**}/J_p$ (where $h^{**}$ is a upper critical field which depends on the temperature), the general configuration is depicted in
figure \ref{fig:spinlayout}(c). The system splits in two sublattices of spin pairs with azimuthal freedom and fixed $z$ projection in each sublattice, such that the sum of the projections satisfies
 $\cos\alpha_1+\cos\alpha_2=\frac{h/J_p}{2}$. We call this phase  ``broken umbrella''.
 The lower panel of figure \ref{fig:mc-MvsH} shows that for $h/J_p>2$ the system chooses one pair of opposite sublattices (in this particular case $A-C$)
to be almost parallel to the magnetic field.

For higher magnetic fields, $h/J_p>h^{**}$, the results shown in the lower panel of figure \ref{fig:mc-MvsH}
are consistent  of the ``broken umbrella'' phase, where all spins have the same $z$ projection. We refer to this phase as the ``umbrella'' phase.

It is important to notice that, as mentioned above,  the configuration (a) in figure \ref{fig:spinlayout}
is only possible for $h/J_p\leq2$, whereas the ``broken umbrella''
is allowed  for all values of the magnetic field below the saturation value.
We have already stated that this system is highly degenerate, and there
is clearly a selection mechanism at play, which selects different configurations depending on the
external magnetic field. We emphasize that this selection is between types of configuration that retain some degeneracy: it is 
not between fixed states, but rather a submanifold of states is selected.  

%


In order to study if the system undergoes a phase transition with   temperature from the paramagnetic phase to the different phases
(corresponding to different regions previously exposed), we  perform Monte Carlo simulations
at a fixed magnetic field using the  simulated annealing technique \cite{Kirkpatrick1983},
lowering the temperature as $T_{n+1}=0.9\times T_{n}$ up to $T/J_{p}=10^{-4}$.

We computed the specific heat per spin $C$ as a function of temperature for different values of the magnetic field:
\ba
C=\frac{\langle E^2\rangle-\langle E\rangle^2}{N\,T^2}.
\label{eq:C}
\ea

 Typical curves for both sectors (low and high field) are shown in figure \ref{fig:CvsT}. There are few features that call our attention:
\begin{figure}[htb]
\vspace{0.5cm}
\begin{centering}
\includegraphics[width=0.8\columnwidth]{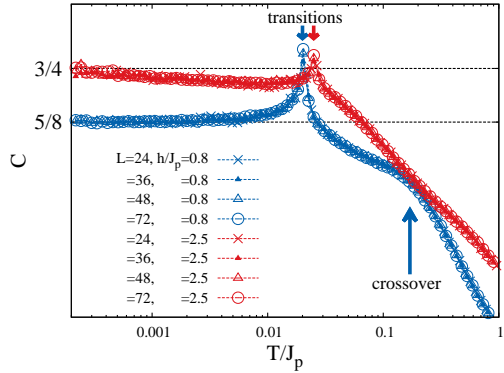}
\end{centering}
\caption{\label{fig:CvsT} Specific heat per spin as a function of temperature for $L=24-72$,  for $h/J_p=0.8,2.5$.
The arrows indicate the limit of the different phases. This transitions point are consistent with the behavior of $|S^z_i|$.}
\end{figure}
\begin{enumerate}
 \item[i-] for $h/J_p<2$ the specific heat shows three different characteristics  in this region (see  figure \ref{fig:CvsT}, blue points).
 The high-temperature regime $0.15\lesssim T/J_p$ corresponds to the paramagnetic phase. In the intermediate regime $0.025\lesssim T/J_p\lesssim 0.15$,
 the internal energy reaches its classical  minimum value $E/N=-\frac{J_p}{2}\left(1+\frac{(h/J_p)^2}{4}\right)$ up to a small contribution from
 thermal fluctuations. Spins on  plaquettes become strongly correlated and satisfy approximately the constraint
 condition $\Sp_{\boxtimes}=\frac{{\bf h}}{J_p}$. This regime is commonly known as a cooperative paramagnet (CP) or classical spin liquid
 \cite{Zhitomirsky2002,Buhrandt2014}.
For $T/J_p\lesssim 0.025$ there is an order-by-disorder selection of the spin configuration shown in figure \ref{fig:spinlayout}(a)  indicated by a reduced specific heat $C=5/8$
\cite{Moessner98,Zhitomirsky2002,Buhrandt2014,Chalker1992,Champion2004,Moliner2009,Sela2014}.
\item[ii-] for $h/J_p>2$  at higher temperatures there is a paramagnetic regime and  for lower the temperatures the spins are
  canted with the magnetic field. In contrast with what happens for $h/J_p<2$ there is not a CP phase
and for  $T/J_p\lesssim 0.03$ there is another order-by-disorder selection with a specific heat $C\simeq 3/4$.
\end{enumerate}

The limiting values of the specific heat at lower temperatures and the evidence of order-by-disorder will be discussed in the following subsection.

Finally, let us notice that the low-temperature phases (figure \ref{fig:spinlayout}) 
break the $\mathcal{Z}_2$  symmetry.
To detect this phase transition we introduce the local order parameter $O_{\mathcal{Z}_2}$ as
\begin{eqnarray}
O_{\mathcal{Z}_2}&=&\frac{4}{N}\sum_{\vr}(\Sp_{\vr,A}+\Sp_{\vr,C}-\Sp_{\vr,B}-\Sp_{\vr,D})
\label{eq:OPz2}
\end{eqnarray}
Following the standard procedure, the second-order transition between the symmetric phase  (large-$T$) and a the broken phase (low-$T$) phase may be located by
the crossing point of the corresponding susceptibility $\chi_{\mathcal{Z}_2}$ and Binder cumulant measured for different clusters.
We have used instantaneous values of equation ~(\ref{eq:OPz2}) to measure the susceptibility $\chi_{\mathcal{Z}_2}$ and 
Binder cumulant $U_{\mathcal{Z}_2}$ associated with this order parameter defined as
\begin{eqnarray}
\chi_{\mathcal{Z}_2}&=&\frac{N_c}{T}\langle(O_{\mathcal{Z}_2})^2\rangle \\
U_{\mathcal{Z}_2}&=&\frac{\langle(O_{\mathcal{Z}_2})^4\rangle}{\langle(O_{\mathcal{Z}_2})^2\rangle^2} 
\end{eqnarray}
\begin{figure}[htb]
\vspace{0.5cm}
\begin{centering}
\includegraphics[width=0.85\columnwidth]{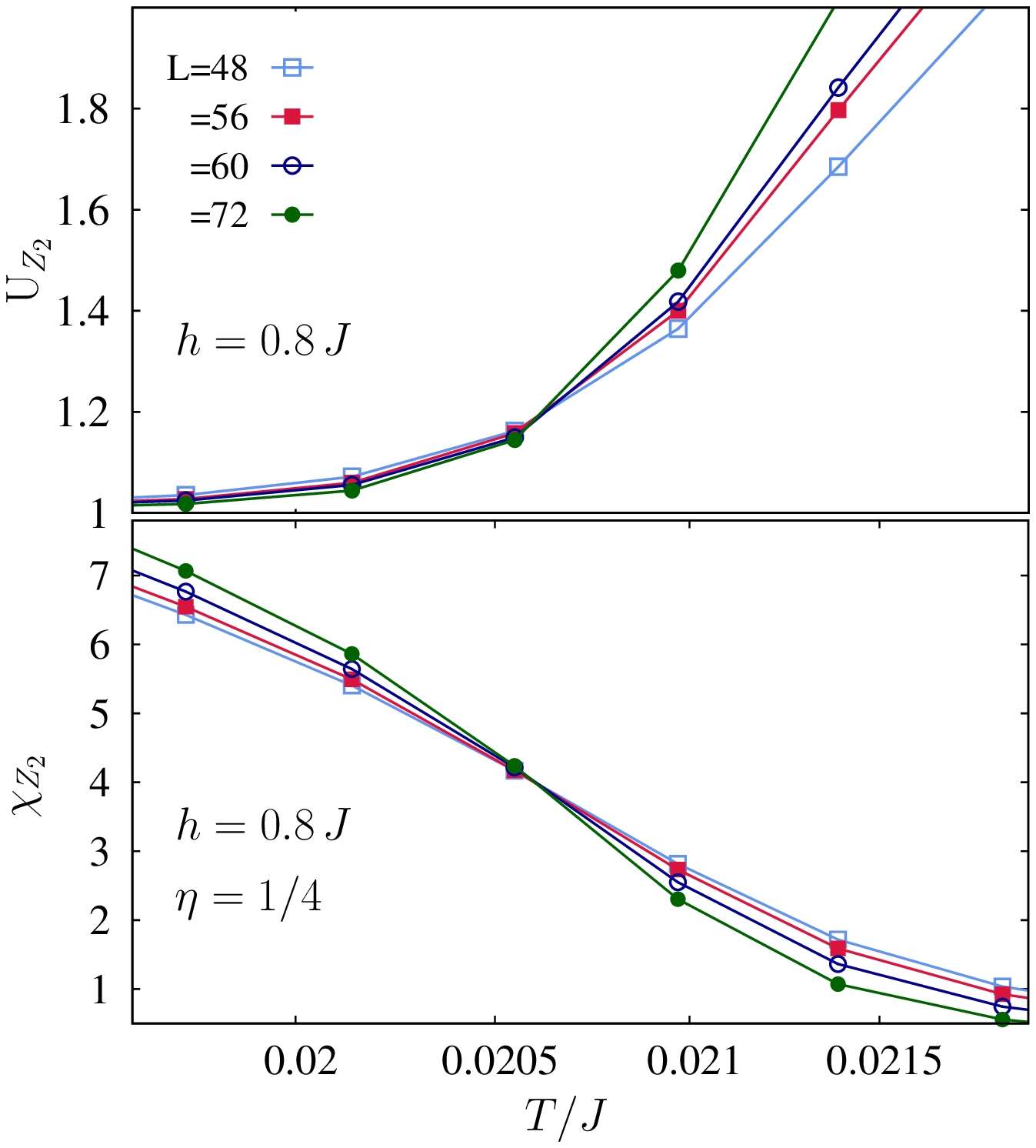}
\includegraphics[width=0.85\columnwidth]{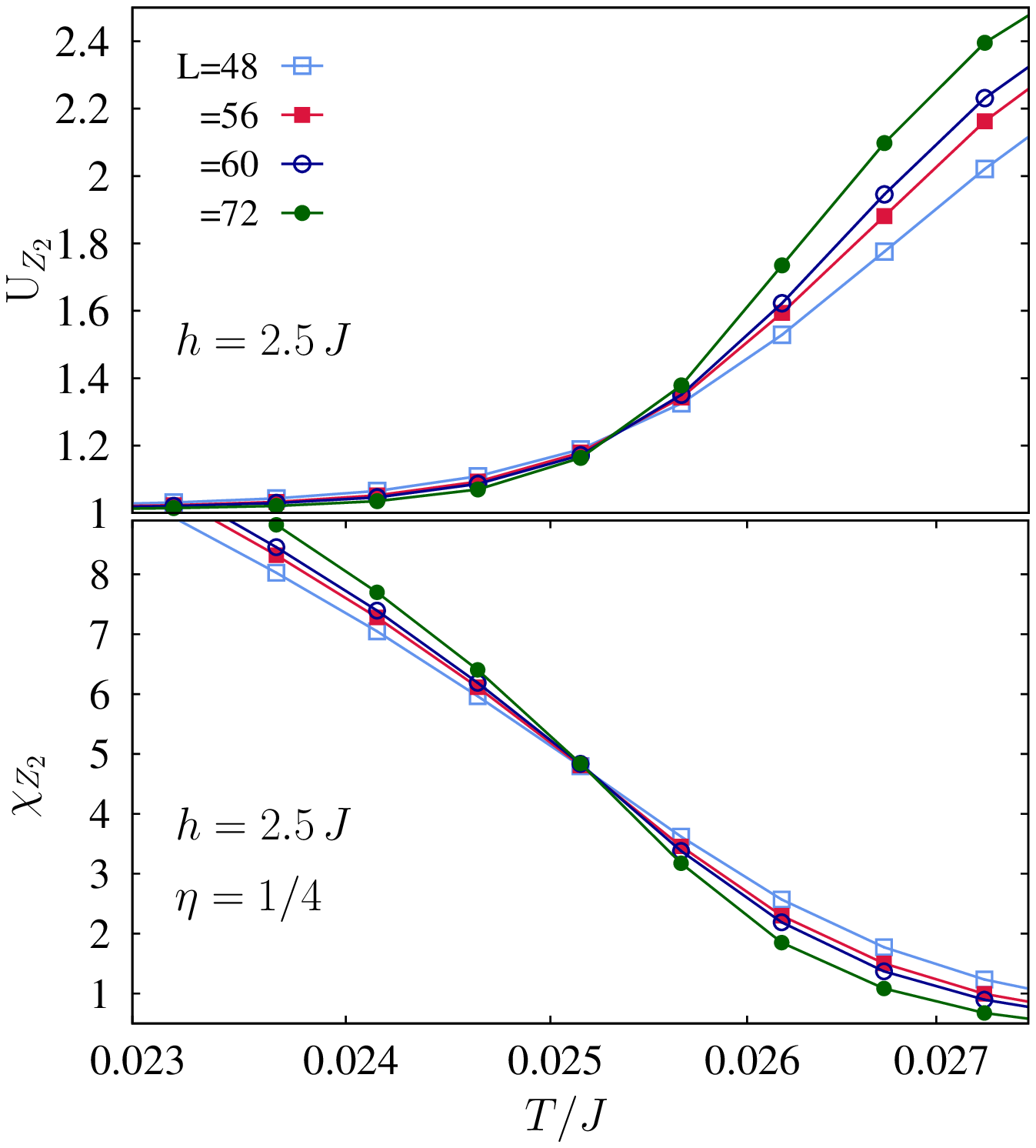}
\end{centering}
\caption{\label{fig:BinderCumul}Transition between the symmetric and broken $\mathcal{Z}_2$ phase in the
at $h/J_p =0.8$ (two upper panels) and $h/J_p =2.5$ (two lower panels). 
Binder cumulant $U_{\mathcal{Z}_2}$ and normalized susceptibility $\chi_{\mathcal{Z}_2}/L^{2-\eta}$ versus temperature for different lattice sizes $L$. 
The error bars for each point,  which are of the order of the point sizes,  were estimated from 20 independent runs
initialized by different random numbers.}
\end{figure}

We illustrate this method in figure  \ref{fig:BinderCumul} for the transition between the symmetric and broken $\mathcal{Z}_2$ phase for 
two magnetic fields: $h/J_p=0.8,2.5$.
Since the critical exponent is known precisely, $\eta=\frac{1}{4}$ \cite{Pathria}, the susceptibility can be studied in the critical region.
In this region, it scales with size $L$  as $\chi_{\mathcal{Z}_2} = L^{2-\eta}f(|1-\frac{T}{T_c}|L^{1/\nu})$. Therefore, 
the normalized susceptibility is size independent at the critical temperature, and curves as a function of $L$ should show a crossing point.
We can observe that the curves for different $L$ plotted as functions of $T$ for both the Binder cumulant
and the normalized susceptibility exhibit a crossing point at the critical temperature.

\subsection{Order-by-disorder selection.}
\label{sec:OBD}
As mentioned earlier, from our simulations we see that  below saturation the specific heat per spin (in units of $k_B$)
is lower than $1$ at the lower temperatures.
A lower value of $C$ is the indication that there are modes that do not
contribute at quadratic order, which suggests the possibility of order-by-disorder selection
\cite{Moessner98,Zhitomirsky2002,Buhrandt2014,Chalker1992,Champion2004,Moliner2009,Sela2014}.
 These modes can be found studying the fluctuation spectrum  up to quadratic order
and looking for zero eigenvalues. Those modes that do not contribute at quadratic order, but they do at quartic order with $\frac{k_BT}{4}$, are called \textit{soft modes}\cite{Moessner98}. There are also modes,
associated with continuous classical degeneracies, that do not contribute at any order:
these are called \textit{zero modes} \cite{Moessner98}.
The distinction whether the zero eigenvalues of the quadratic fluctuations
are either soft or zero modes can be done inspecting the $C$ curves at low temperatures \cite{Moessner98}.

Taking in consideration the discussion above, we aim to study the low temperature 
selection between configurations in our system shown in figure \ref{fig:spinlayout}.
This selection \textit{partially} lifts the degeneracy of the system, and a group of ground states with continuous degeneracies is chosen depending on
the external  magnetic field. 
We propose that this partial selection is due to the thermal order by disorder mechanism. In order
to study this, we calculate the spectrum of the quadratic fluctuations \cite{Moessner98,Zhitomirsky2002,Buhrandt2014,Chalker1992,Champion2004,Moliner2009,Sela2014}.
 for each selected groundstate manifold. If zero eigenvalues are obtained,
following \cite{Moessner98}, we inspect the specific heat curves obtained from the simulations to determine whether these are soft or zero modes.

We calculated the energy per unit cell for the configurations found  (a), (c) from figure \ref{fig:spinlayout} and
introduced quadratic fluctuations for each configuration with each spin parametrized (in their own frame) as
\begin{equation}
\Sp_{\vr,\mu}=\left(\epsilon^x_\mu(\vr),\, \epsilon^y_\mu(\vr),\, \epsilon^z_\mu(\vr) \right)
\end{equation}
where $\mu=A,B,C,D$, $\epsilon^z_\mu(\vr) = 1-\frac{1}{2}\left(\epsilon^x_\mu(\vr)^2 + \epsilon^y_\mu(\vr)^2 \right)$,
which verifies $|\Sp_{\vr,\mu}|^2=1$ up to quadratic order.
We then construct the fluctuation matrix in the Fourier space for a unit cell of each configuration,
thus getting in each case a $8 \times 8$ matrix  which can be diagonalized analytically (explicit expressions are given in Appendix A).

In the case of \ref{fig:spinlayout}(a),``the dimer phase'', there are three continuous degeneracies  associated  with the ``AF dimers'' in one sublattice, and the
local azimuthal rotation of each of the pairs in the other sublattice.
If we consider these associated to the three zero eigenvalues found by the fluctuation analysis, thus having three zero modes,
we get $C=\frac{1}{4}\left(3\cdot0 + 5\cdot\frac{1}{2}\right)=\frac{5}{8}=0.625$, which is the low temperature value of $C$ obtained in the simulations.
A larger $C$ is expected if either one of them is a soft mode. Notice that in section \ref{sec-model}, this configuration was compared to other possible ones.
Four different possible arrangements of spins per plaquette, including this one, were shown in figure \ref{fig:Ej-spin-conf}. Of the four cases discussed, 
this one is the one wich retains more degrees of freedom. 
At low enough temperatures, by the order-by-disorder mechanism, the system chooses this configuration because it has more degrees of freedom, which are reflected as  
 more zero modes.

In the particular case of $h/J_p=2$ (configuration \ref{fig:spinlayout}(b))
there are only two continuous degeneracies and four zero eigenvalues.
We take two of them to be zero modes and two to be soft modes,
 and $C=\frac{1}{4}\left(2\cdot0 +  2\cdot\frac{1}{4} + 4\cdot\frac{1}{2}\right)=\frac{5}{8}=0.625$.
As in the ``dimer phase'' case, this mode counting reproduces the results from the simulations. 
 
We repeat the procedure for the high field region. 
Configurations like figure  \ref{fig:spinlayout}(c) have two zero eigenvalues and there are two continuous  degeneracies.
If the two zero eigenvalues are zero modes, then again $C= \frac{1}{4}\left(2\cdot0 + 6\cdot\frac{1}{2}\right)=\frac{3}{4} = 0.75$.

\begin{figure}[htb]
\begin{centering}
\includegraphics[width=0.9\columnwidth]{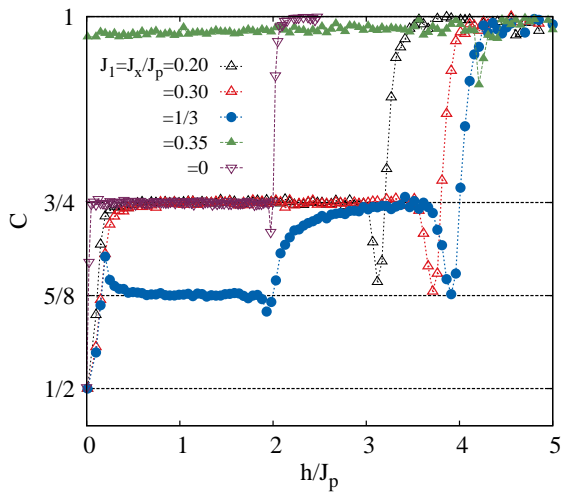}
\par\end{centering}
\caption{\label{fig:cvsh} Specific heat as a function of the external magnetic field for $J_1=J_x=0.2J_p,\, 0.3J_p,\,J_p/3,0.35J_p$
obtained from Monte-Carlo simulations  at $T/J_p=5\times10^{-4}$, $L=72$. Lines are drawn
at $C=\frac{5}{8}(0.625),\frac{3}{4}(0.75)$.
To compare with the case of isolated dimers we shown the $J_1=J_x=0$ curve. We observe that for $J_1=J_x<J_p/3$ all the curves have similar behavior.}
\end{figure}

The specific heat obtained from the simulations as a function of temperature was shown in figure \ref{fig:CvsT} for two values of the magnetic field.
Figure \ref{fig:cvsh} shows the specific heat at  $T/J_p=1\times10^{-3}$ as a function of the magnetic field for
some values of the ratios $J_1/J_p,J_x/J_p$ (in the next section we will discuss the system behavior outside the point $J_1=J_x=J_p/3$).
Dashed horizontal lines  are drawn at $C=\frac{5}{8}(0.625)$ and $\frac{3}{4}(0.75)$, the low field and high field values, respectively.

For the lowest magnetic field,  the specific heat per spin is close to $1/2$.
Fluctuations for a planar configuration where the systems splits in two independent ``AF dimer'' sublattices
have four zero eigenvalues, which gives $C=\frac{1}{2}$ when considered as zero modes. If any of these values were a soft mode, 
a larger $C$ would be obtained.

The fluctuation  and specific heat analysis allows us to conclude that at $J_1=J_x=J_p/3$
the configurations selection along the whole magnetization curve is due to the order-by-disorder mechanism.
In particular, the change of behavior of the system at $h/J_p=2$ is also due to this mechanism:
for  $h/J_p<2$ there is a type of possible configuration that has more zero modes, and is thus selected.

In the case of $h/J_p>2$, the system tends towards a particular case, where two sublattices were completely
aligned with the magnetic field.
To check this, we performed the spin-wave analysis similar
to one proposed by Kawamura  \cite{Kawamura1984}
for a general ``broken umbrella'' layout (see figure \ref{fig:spinlayout}(c)).
We kept those solutions that were physically relevant for the problem, i.e. that existed
for $2<h/J_p<4$ and that implied $0\leq\alpha_{1,2}<\frac{\pi}{2}$. The minimum was found for $\alpha_1=0,\alpha_2=\frac{h}{2}-1$.
We conclude that in the limit of $T\to 0^+$ this particular case of the ``broken umbrella'' is selected.

\begin{figure}[htb]
\vspace{0.5cm}
\begin{centering}
\includegraphics[width=0.98\columnwidth]{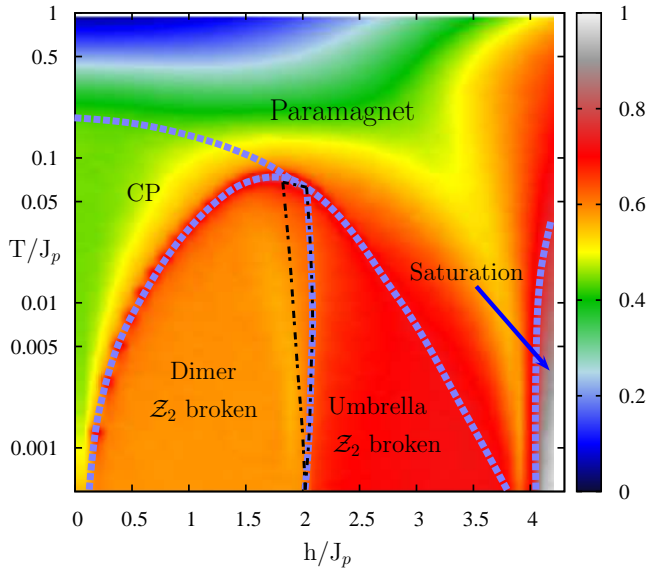}
\par\end{centering}
\caption{\label{fig:phase-diagram}(Color online) Phase diagram of the model given by equation \ref{eq:Hamil} in the strongly frustrated point $J_1=J_x=J_p/3$.
At higher temperatures, there is a crossover
from a conventional  paramagnetic phase to a classical spin-liquid (CSL) or cooperative paramagnet. At lower $T$, for  $h/J_p<2$ there is an
order-by-disorder selection to a ``dimer'' phase (see figure \ref{fig:spinlayout}a ).
Around $h/J_p=2$ at low temperature there is  a pseudo plateau at $M=1/2$ with configuration (b) in figure .\ref{fig:spinlayout}. 
This is indicated in the area surrounded by a thin black dashed line.
For $h/J_p>2$ coming from the CSL phase there is first an order-by-
disorder selection to a  broken $\mathcal{Z}_2$ phase with   spin configuration (c) of figure \ref{fig:spinlayout}c ). In the limit $T\to 0$, for  $h/J_p>2$
there is a new order-by-disorder selection inside the manifold of ground states given by figure \ref{fig:spinlayout}c. }
\end{figure}

Having discussed the different spin configurations at the lowest temperatures, the selection mechanisms and the specific heat, we can present
the main results of our study in the phase diagram in  figure \ref{fig:phase-diagram}.
As a background we show the density plot of the specific heat vs temperature and magnetic field and over this, lines indicating the edges of the different phases.
The pseudo plateau region, obtained studying the susceptibility, is also shown.
Coming from the high temperature region, we have a crossover from a conventional  paramagnetic phase to a cooperative paramagnet.

The edge between the conventional  paramagnetic phase to a CP phase is indicated by a crossover  in the specific heat curve. (see figure \ref{fig:CvsT}).

Lowering the temperature even more, for the low field region there is a second order transition corresponding to an order-by-disorder selection to a phase with
broken $\mathcal{Z}_2$  symmetry. 
This ``dimer phase'' consists of two sublattices of spin pairs, where one pair satisfies the groundstate plaquette condition and
the other one is free (``AF dimer''). The transition between the CP and the dimer phase is manifested by a peak in the specific heat
and by the point where the Binder cumulant  and the order parameter susceptibility-curves cross  for different sizes .

Around the special value of the magnetic field $h/J_p=2$, at low temperatures a not fully collinear pseudo plateau stabilizes at $M=1/2$
where there is an alternation of  pairs of spins polarized with the magnetic field and pairs in the ``AF dimer'' state. 

For higher magnetic fields, coming from the paramagnetic phase we have an order-by-disorder selection to a broken $\mathcal{Z}_2$ phase with a ``broken umbrella like''
configuration.
As in the low field region, this transition is manifested by a peak in the specific heat  and by the point where the Binder
cumulant and the order parameter susceptibility-curves cross  for different sizes.
In the limit $T\to 0$ and  $h/J_p>2$ there is a new order-by-disorder selection inside the manifold of ground states,
where a pair of spins is completely aligned with the magnetic field.

\section{Weakly frustrated region $J_1=J_x<J_p/3$}
\label{sec-frustrated-line}

In this section, we extend the above analysis to $J_1=J_x<J_p/3$.
The spherical approximation suggested that for this range the ground state is degenerate.
This range of couplings corresponds to the Hamiltonian presented in equation \ref{eq:Hpairs}, which is valid along the whole line $J_1=J_x$ in parameter space.
Minimizing with respect to the total spin of the opposite pairs per plaquette, $\mathbf{P}_{\boxtimes,i}$ we obtain the ground state condition:

\begin{eqnarray}
\frac{J_p}{3}\mathbf{P}_{\boxtimes,\alpha} + J_1\mathbf{P}_{\boxtimes,\beta} - \frac{\mathbf{h}}{3} = 0 \\
\label{eq:gsweak1}
\frac{J_p}{3}\mathbf{P}_{\boxtimes,\beta} + J_1\mathbf{P}_{\boxtimes,\alpha} - \frac{\mathbf{h}}{3} = 0
\label{eq:gsweak2}
\end{eqnarray} 

We focus on the solutions for $J_1\neq J_p/3$, since the highly frustrated point has been discussed in the previous section.
These are:

\begin{eqnarray}
\label{eq:gszweak}
\mathbf{P}_{\boxtimes,i}^z &=& \frac{\mathbf{h}}{J_p+3J_1} \\
\label{eq:gsxyweak1}
\mathbf{P}_{\boxtimes,\alpha}^{xy}&=&\mathbf{P}_{\boxtimes,\beta}^{xy}=0 \\
\,\,\,&\text{or}&\,\,\, \nonumber \\ 
\mathbf{P}_{\boxtimes,\alpha}^{xy}&=&-\mathbf{P}_{\boxtimes,\beta}^{xy};\,|\mathbf{P}_{\boxtimes,\alpha}^{xy}|=\sqrt{4-(\mathbf{P}_{\boxtimes,i}^z)^2}
\label{eq:gsxyweak2}
\end{eqnarray}

\noindent where we have decomposed each spin pair vector in the projection along and perpendicular to the magnetic field:
$\mathbf{P}_{\boxtimes,i}=\mathbf{P}_{\boxtimes,i}^z+\mathbf{P}_{\boxtimes,i}^{xy}$.

Evaluating the energy, it can be seen that for $J_1=J_x <J_p/3$ the solution for the projection of the spin dimers perpendicular 
to the external magnetic field is that on equation (\ref{eq:gsxyweak1}), which is that each pair has a zero projection in the $xy$ plane.
This implies that in each spin pair the components parallel to the external field are fixed, but those in the ortogonal plane cancel out and thus there is an azimuthal
freedom in each pair. This configuration is the one illustrated in figure \ref{fig:Ej-spin-conf}(a). 
For  $J_1=J_x > J_p/3$, the lowest energy is found for fixed components, equation (\ref{eq:gsxyweak2}),
as shown in figure \ref{fig:Ej-spin-conf}(b).

\begin{figure}[htb]
\begin{centering}
\includegraphics[width=1\columnwidth]{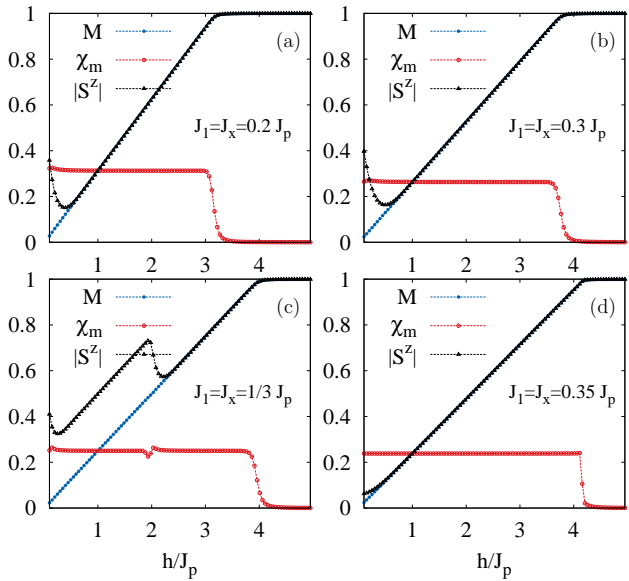}
\end{centering}
\caption{\label{fig:mag_moreJ} Magnetization curves for $J_1=J_x=0.2J_p,\, 0.3J_p,\,J_p/3,0.35J_p$ obtained from Monte-Carlo simulations for $L=72$,
$T/J_p=1\cdot10^{-3}$.
Magnetization, absolute value of the magnetization
and susceptibility are shown in each case. It can be seen that the only distinct behavior is that of the highly frustrated point $J_1=J_x=J_p/3$.}
\end{figure}

\begin{figure}[htb]
\begin{centering}
\includegraphics[width=1\columnwidth]{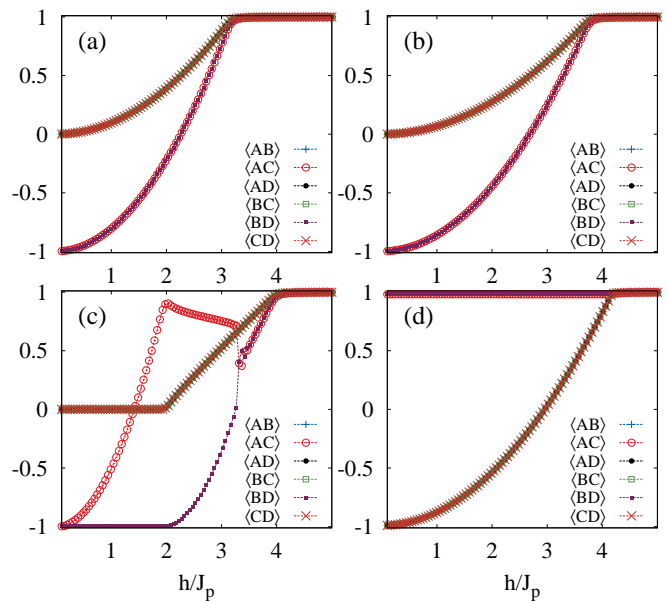}
\end{centering}
\caption{\label{fig:scalar_moreJ} Average scalar product $\langle\Sp_{\vr,i}\cdot\Sp_{\vr,j}\rangle=\langle ij\rangle$  curves  ($i,j=A,B,C,D$ in the legend)
for $J_1/J_p=J_x/J_p=0.2$ (a), $0.3$ (b), $1/3$ (c) and $0.35$ (d) obtained from Monte-Carlo simulations for $L=72$.}
\end{figure}

Following the discussion for the highly frustrated point, through simulations we study the magnetizations curves, the scalar products and the specific heat for
different values of the couplings along the $J_1=J_x$ line to see if there is a state selection at low temperatures.  
In figure ~\ref{fig:mag_moreJ} we show the magnetization curves for $J_1=J_x=0.2J_p$ and $J_1=J_x=0.3J_p$ compared with $J_1=J_x=J_p/3$ and
with a higher value $J_1=J_x=0.35J_p > J_p/3$.
We see that there is no particular feature in the curves for any value of the couplings, besides the pseudo plateau at $J_1=J_x=J_p/3$ that
has been shown and discussed in the previous section.
  Results for the average scalar
product $\langle\Sp_{\vr,i}\cdot\Sp_{\vr,j}\rangle$ are shown in figure \ref{fig:scalar_moreJ}
for two values that satisfy $J_1=J_x<J_p/3$ (a),(b) to be compared with  $J_1=J_x=J_p/3$ (c) and  $J_1=J_x=0.35J_p>J_p/3$  (d).
The behaviour matches the calculations described above.
The spins all have the same $S_z$ component, thus the system behaves effectively as an $xy$ model with zero magnetization between opposite pairs.
There is no particular selection among these configurations, up to the temperatures presented in this work.
In the $J_1=J_x>J_p/3$ case, spins connected by $J_p$ are parallel along all the magnetization curve.
This phase is simply an effective planar configuration  where each layer is a $xy$-Neel.


The configurations for the low frustrated case have an azimuthal degree of freedom that should be seen as zero modes in the specific heat.
Example curves for $L=72$ at $T/J_p=10^{-3}$,  $J_1=J_x=0.2J_p,\,0.3J_p$ are shown in
figure \ref{fig:cvsh}. Results for $J_1=J_x=J_p/3$ and $J_1=J_x=0.35J_p>J_p/3$ are also shown  for comparison.
The specific heat for the lower values of the couplings is $C \sim 3/4$,
which agrees with that of the strongly frustrated point at the higher magnetic fields for the strongly frustrated point.
However, in this case there is no order by disorder \cite{Moessner98b}. 
This decrease in the specific heat is due to the presence of two zero modes  in the ground state 
(continuous degeneracies associated with azimuthal rotation of each pair joined by $J_p$).
For comparison, the specific heat for $J_1=J_x=0.35J_p>J_p/3$ is $C \sim 1$, which implies, that there are no zero nor soft modes. Quadratic fluctuations
have no zero eigenvalues.

Finally, in the figure \ref{fig:cvsh} we also show the specific heat curve for the case of isolated classical dimers ($J_1=J_x=0$).
Inspection of the specific heat and the scalar products between (figure \ref{fig:scalar_moreJ}) for $J_1=J_x<J_p/3$,
show a remarkable similarity to this extreme case, suggesting that in all the range $J_1=J_x<J_p/3$ the system behaves as a classical dimer lattice.

\section{Conclusions}
\label{sec-conclusions}
%

In summary, we have studied the phase diagram of the classical Heisenberg $J_1$-$J_x$-$J_p$ antiferromagnetic model
on a bilayer honeycomb lattice in a magnetic field.
Using a combination of analytical considerations and classical Monte Carlo,
we have found a very rich low temperature phase diagram for the strongly frustrated point $J_p/3=J_1=J_x$,
which is summarized in Fig.~\ref{fig:phase-diagram}.
The phase diagram features three nontrivial regions characterized by broken lattice symmetries and
different entropic order by-disorder selection.
These selections partially lift the ground state degeneracy, since the chosen phases retain different degrees of freedom, wich contribute with 
zero and soft modes, for different values  of the external magnetic field.

Coming from the high temperature region we observe a crossover from a conventional  paramagnetic phase to a
cooperative paramagnet (CP). The boundary between the paramagnetic phase to a CP is shown by a smooth but
a clear crossover in the specific heat, see figure \ref{fig:CvsT}.
Lowering the temperature even more, in the low field region $h/J_p<2$ there is an order-by-disorder selection
to a phase with broken $\mathcal{Z}_2$.
In this ``dimer phase'' one couple of opposite spins  is canted
with the magnetic field and the other pair is antiparallel and random corresponding to a kind of ``free AF dimer state''
(see figure \ref{fig:spinlayout}(a)). The transition between the CP
and the dimer phase is manifested by a peak in the specific heat (see figure \ref{fig:CvsT}).
By the computation of the Binder cumulant and the order parameter susceptibility
we have  numerically checked that this is a second order phase transition where $\mathcal{Z}_2$ is broken.

Around $h/J_p=2$, a not fully collinear  classical 1/2-magnetization plateau is stabilized at low temperature. 
This new kind of classical plateau is in a strong contrast to the fully collinear classical plateau states in the frustrated  lattice 
as the triangular\cite{Kawamura1984} and honeycomb\cite{Rosales2013} cases.
This remarkable property is due to the very large  configurational degeneracy of the system of frustrated ``classical spin dimers''.
The stabilized structure (figure ~\ref{fig:spinlayout}(b)), consists of alternation of polarized and anti-parallel spin pairs in 
the plateau phase which breaks  the $\mathcal{Z}_2$ symmetry between two sites in a unit cell of the honeycomb lattice.

For $h/J_p>2$ coming from the cooperative paramagnet phase, we have  first an order-by-disorder selection
to a broken $\mathcal{Z}_2$ phase with a ``broken umbrella'' configuration (see figure \ref{fig:spinlayout}(c) ),
 which was  confirmed by the analysis of the order parameter susceptibility and the Binder cumulant as a continuous transition.
In the limit $T\to 0$ and  $h/J_p>2$ there is a new order-by-disorder selection inside the manifold of ground states given by figure \ref{fig:spinlayout}(c),
where a pair of spins is completely aligned with the magnetic field.

Finally, along the $J_1=J_x$ line the system can be effectively described as  a function of opposite spin pairs. 
The configurations for $J_1=J_x>J_p/3$ are fixed, whereas for the less frustrated region  $J_1=J_x<J_p/3$ the ground state is degenerate.
The system behaves as an effective
$xy$ model and retains an azimuthal degeneracy in opposite spin pairs.
These zero modes give a lower specific heat $C=3/4$, but there is no order by disorder.

We mention that the present work  may be relevant in the study of different compounds
that are described by the frustrated hexagonal Heisenberg model, such as Bi$_3$Mn$_4$O$_{12}$(NO$_3$) \cite{Smirnova2009} with spin-$3/2$.

\section*{Acknowledgments}
The authors specially thank  M. Zhitomirsky for fruitful discussions and careful reading of the manuscript,
and  D. Cabra and  R. Borzi for communicating and discussing  results.
This work was partially supported by CONICET (PIP 0747) and ANPCyT (PICT 2012-1724).

\bigskip
\appendix
\section{Fluctuation matrices}
\label{AppendixA}

We analize the fluctuations for each configuration found in our simulations, (a), (c)
from figure \ref{fig:spinlayout} (the configuration (b) it’s just a special case of (a)). We parametrize the spins with the same angles
as in figure \ref{fig:spinlayout}, so that 
\begin{eqnarray}
\Sp_{\vr,\nu}^{(a)}&=&\{\sigma\cos\beta_{1,\vr}\sin\alpha_1,\sigma\sin\beta_{1,\vr}\sin\alpha_1,\cos\alpha_1 \} \nonumber \\
\Sp_{\vr,\mu}^{(a)}&=&\sigma\{\cos\beta_{2,\vr}\sin\alpha_2,\sin\beta_{2,\vr}\sin\alpha_2,\cos\alpha_2 \}\nonumber \\
\Sp_{\vr,\nu}^{(c)}&=&\{\sigma\cos\beta_{1,\vr}\sin\alpha_1,\sigma\sin\beta_{1,\vr}\sin\alpha_1,\cos\alpha_1 \} \nonumber \\
\Sp_{\vr,\mu}^{(c)}&=&\{\sigma\cos\beta_{2,\vr}\sin\alpha_2,\sigma\sin\beta_{2,\vr}\sin\alpha_2,\cos\alpha_2 \} \nonumber\\
\label{eq:Sp_parametri}
\end{eqnarray}
where $\beta_{i,\vr}$ can be different in each plaquette, $\nu=A,C$, $\mu=B,D$ and $\sigma=+$ for
sites $A,B$ and $\sigma=-$ for $C,D$. Under fluctuations their new coordinates are expressed in their own frame as:
\begin{eqnarray}
\Sp_{\vr,\nu}=\{\epsilon^{x}_{\vr,\nu},\epsilon^{y}_{\vr,\nu},1-\frac{1}{2}\left[\left(\epsilon^{x}_{\vr,\nu}\right)^2+\left(\epsilon^{y}_{\vr,\nu}\right)^2\right]\}
\end{eqnarray}
is verifying the condition $|\Sp_{\vr,\nu}|^2=1$ up to quadratic order. Following Ref. \cite{Chalker1992}  the Hamiltonian, in the Fourier space, is expanded up to second order in spin deviations as:
\begin{eqnarray}
\mathcal{H}&=&E_{0}+\sum_{\vk}\Psi(-\vk)\cdot\mathcal{M}\cdot\Psi(\vk)
\label{eq:H-OBD-appendix}
\end{eqnarray}

where the vectors $\Psi(\vk)$ read
\begin{eqnarray}
\Psi(\vk)=\{\epsilon^{x}_{\vk,A},\epsilon^{y}_{\vk,A},\epsilon^{x}_{\vk,C},\epsilon^{y}_{\vk,C},\epsilon^{x}_{\vk,B},\epsilon^{y}_{\vk,B},\epsilon^{x}_{\vk,D},\epsilon^{y}_{\vk,D}\} \nonumber
\end{eqnarray}
In Eq. (\ref{eq:H-OBD-appendix}) the $8\times8$ matrix $\mathcal{M}$ can be written as 
\begin{equation}
\mathcal{M}=\left[ \begin{array}{cc}
m_{11} & m_{12} \\
m^{\dagger}_{12} & m_{22}
\end{array} \right].
\label{eq:OBDmatrix1}
\end{equation}
where the complex $4\times4$ matrices $m_{ij}$ depend of the angles in the spin configuration.

In the low magnetic field region (configuration (a)), the matrices $m_{ij}$  can be written in a compact 
form using the Kronecker product as
\begin{eqnarray}
m_{11}&=&\sigma_0\otimes\sigma_0+\sigma_x\otimes(\cos(\alpha)^2\,\sigma_0-\sin(\alpha)^2\sigma_z)\\
m_{22}&=&\sigma_0\otimes\sigma_0-\sigma_x\otimes\sigma_z\\
m_{12}&=&\frac{\gamma_{\vk}}{3}\left[ \begin{array}{cccc}
c_{\alpha\beta}\,c_{\gamma}+s_{\alpha\beta}  &  -c_{\alpha}s_{\gamma}  &  -c_{\alpha\beta}c_{\gamma}-s_{\alpha\beta} & -c_{\alpha}s_{\gamma}\\
c_{\beta}s_{\gamma}  &  c_{\gamma}  &  -c_{\beta}s_{\gamma}  &  c_{\gamma}\\
c_{\alpha\beta}c_{\gamma}-s_{\alpha\beta}  &-  c_{\alpha}s_{\gamma}  &  -c_{\alpha\beta}c_{\gamma}+s_{\alpha\beta} & -c_{\alpha}s_{\gamma}\\
c_{\beta}s_{\gamma}  &  c_{\gamma}  &  -c_{\beta}s_{\gamma}  &  c_{\gamma}
\end{array} \right]\nn\\
\end{eqnarray}

where $\sigma_0$ is the $2\times2$ identity matrix and $\sigma_a$ ($a=x,y,z$) are the Pauli's matrices; 
$c_{\alpha\beta}=\cos\alpha\cos\beta$, $s_{\alpha\beta}=\sin\alpha\sin\beta$, $c_\alpha=\cos\alpha$, $s_\alpha=\sin\alpha$  (same for $\beta,\gamma$). 

In the high field region  (configuration (c)), we have

\begin{eqnarray}
m_{11}&=&\sigma_0\otimes\sigma_0+\sigma_x \otimes \sigma_0\\
m_{22}&=&\sigma_0\otimes\sigma_0+\sigma_x\otimes(\cos(\beta)^2\,\sigma_0-\sin(\beta)^2\sigma_z)\\
m_{12}&=&\frac{\gamma_{\vk}}{3}\left[ \begin{array}{cccc}
c_\beta c_\gamma & -s_\gamma& c_\beta c_\gamma & -s_\gamma\\
c_\beta s_\gamma & c_\gamma& c_\beta s_\gamma & c_\gamma\\
c_\beta c_\gamma & -s_\gamma& c_\beta c_\gamma & -s_\gamma\\
c_\beta s_\gamma & c_\gamma& c_\beta s_\gamma & c_\gamma
\end{array} \right].\nn\\
\end{eqnarray}

\end{document}